\documentclass[12pt]{iopart}
\usepackage{amssymb}
\expandafter\let\csname equation*\endcsname\relax
\expandafter\let\csname endequation*\endcsname\relax
\usepackage{amsmath}
\usepackage{graphicx}
\usepackage{color}
\usepackage{txfonts}
\usepackage{microtype}
\usepackage[normalem]{ulem}
\usepackage[english]{babel}
\usepackage[T1]{fontenc}

\newcommand{\beq}{\begin{equation}}
\newcommand{\eeq}{\end{equation}}
\newcommand{\beqa}{\begin{eqnarray}}
\newcommand{\eeqa}{\end{eqnarray}}

\begin{document}

\title{Detecting composite orders in layered models via machine learning}

\author{W.~Rz\k{a}dkowski$^1$, N.~Defenu$^2$, S.~Chiacchiera$^3$, A.~Trombettoni$^{4,5}$ and G.Bighin$^1$}

\address{$^1$Institute of Science and Technology Austria (IST Austria), Am Campus 1, 3400 Klosterneuburg, Austria \newline
$^2$Institut f\"ur Theoretische Physik, Universit\"at Heidelberg, D-69120 Heidelberg, Germany\newline
$^3$Science and Technology Facilities Council (STFC/UKRI), Daresbury Laboratory, Keckwick Lane, Daresbury, Warrington WA44AD, United Kingdom \newline
$^4$CNR-IOM DEMOCRITOS Simulation Center, Via Bonomea 265, I-34136 Trieste, Italy \newline
$^5$SISSA and INFN, Sezione di Trieste, Via Bonomea 265, I-34136 Trieste, Italy \newline }

%
%
%

\date{\today}

\begin{abstract}
  Determining the phase diagram of systems consisting of smaller subsystems 'connected' via a tunable coupling is a challenging task relevant for a variety of physical settings. 
    A general question is whether new phases, not present in the uncoupled limit, may arise.
  We use machine learning to study layered spin models,  in which the spin variables constituting each of the uncoupled systems (to which we refer as layers) are coupled to each other via an interlayer coupling. In such systems, in general, composite order parameters involving spins of different layers may emerge as a consequence of
  the interlayer coupling. We focus on the layered Ising and Ashkin-Teller models as a paradigmatic case study,
  determining their phase diagram via the application of a machine learning
  algorithm to the Monte Carlo data. Remarkably our technique is able to correctly characterize all the
  system phases also in the case of hidden order parameters, {\emph{i.e.,}}~order parameters 
  whose expression in terms of the microscopic configurations would require additional preprocessing of the data fed to the algorithm.
  We correctly retrieve the three known phases of the Ashkin-Teller model with ferromagnetic couplings, including the phase described by a composite order parameter.
    For the bilayer and trilayer Ising models the phases we find
    are only the ferromagnetic and the paramagnetic ones. Within the approach we introduce, owing to the construction  of convolutional neural networks, naturally suitable for layered image-like data with arbitrary number of layers, no preprocessing of the Monte Carlo data is needed, also with regard to its spatial structure. The physical meaning of our results is
  discussed and compared with analytical data, where available.
  Yet, the method can be used without any {\emph{a priori}} knowledge
  of the phases one seeks to find and can be applied to other models and structures.
\end{abstract}

\maketitle

\section{Introduction}
Classification of observations into separate categories is certainly one of the most important
applications of machine learning \cite{Carleo:2019us}. Successful examples range broadly from the detection of exotic
particles in experimental high energy physics~\cite{ExoticParticles} through learning human actions
in movies~\cite{Movies} to dermatologist-grade skin cancer classification~\cite{SkinCancer}.
The classification task is very often performed with artificial neural networks, capable of learning
even highly complex and elusive patterns in the data, both detectable and invisible to humans.
If multilayer image data is in question, convolutional neural networks (CNNs) perform exceptionally well,
mimicking human vision by inferring from small portions of the image at a time. Successful applications of CNNs outside the field of physics are extremely numerous, ranging from the detection of human faces at multiple angles or in partially visible images~\cite{HumanFaces} to the ImageNet large-scale classification
  challenge~\cite{ImageNet}, just to mention a few.

Translational invariance and adjustable size of the filters, which detect local correlations,  make CNNs the ideal candidates for phase diagram reconstruction. The phase diagram is typically reconstructed from a large number of Monte Carlo (MC) snapshots. At first,
research efforts revolved around supervised learning on the MC
snapshots~\cite{Carrasquilla:2017iz, vanNieuwenburg:2017hp, Kim:2018kv, Casert:2019ca, Zhao:2018wi, RichterLaskowska:2018co, Dong:2019ed, Zhang:2019di, Beach:2018hs}, later shifting to fully unsupervised learning on a chosen observable,
such as non-local correlators whose behaviour is modified by the presence of phase transitions~\cite{Broecker:2017vva}.

Our goal is to use machine learning techniques to characterize the phase diagram
  of coupled spin models. We refer to them as {\emph{layered}} spin models since it is natural to think of the coupling between two-dimensional spin models considering them separately, say in different layers and involving different spin variables, subsequently turning on the interlayer coupling. The simplest structure of this kind is a bilayer. It is clear
  that, depending on the form and the strength of the coupling, the phases of the uncoupled models can be altered and new phases, impossible without the interlayer coupling, {\emph{may}} emerge.
Therefore, it becomes important to devise a general approach that can detect the presence of phases induced by the interlayer coupling.

In this paper we introduce a CNN-based approach capable of fully unsupervised
learning of phase diagrams with the network fed exclusively with raw MC snapshots without
any {\emph{a priori}} knowledge about relevant observables or order parameters. We note that complementary approaches to the unsupervised learning problem have been pursued using principal component analysis and
support vector machines~\cite{PCA1,PCA2,PCACondensed,Kernel,giannetti2019machine,Comparing}, deep autoencoders~\cite{DeepAutoencoder} or
discriminative cooperative networks~\cite{Liu:2018bv}. 
However, here we show that the task of fully-unsupervised phase diagram
reconstruction can also be performed using CNNs, allowing one to apply to physical problems a number of techniques
developed in the field of computer vision, a field in which CNNs represents the golden standard.

Our approach is applied to the reconstruction of the phase diagram of layered spin models. Our motivation
for such an investigation is three-fold.
On one side, when two or more models are coupled, new phases may emerge
as a result of the presence and of the form of the coupling. Consider, for instance, two magnetic systems with a tunable coupling
between each other and suppose that when the coupling is zero, each system separately undergoes a conventional ferromagnetic phase transition~\cite{Cardy96}.
For finite coupling, on the other hand, the order parameter may involve, in the general case, some non-trivial combination of spins of both systems.
Let us consider a specific example, i.e.~the Ashkin-Teller model, consisting of two square-lattice Ising models with spin variables $\sigma$ and $\tau$ coupled via a term
of the form $\sigma \sigma \tau \tau$~\cite{Ashkin43,Baxter:2007}. When the interlayer coupling between the variables $\sigma$ and $\tau$ is zero, the phase diagram of the model is characterized only by the order parameters $\langle \sigma \rangle$ and $\langle \tau \rangle$. On the other hand,
when the interlayer coupling is large enough with respect to the intralayer term, a new non-trivial phase with a
composite order parameter $\langle \sigma \tau \rangle$ emerges, even when all couplings are ferromagnetic. Further examples of the occurrence
  of novel order due to the coupling between different layers include the so-called `metallic superfluid' phase \cite{Babaev:2004el,Svistunov:2015}, as well as the recently-reported
  BKT-paired phase in two coupled two-dimensional XY models \cite{Bighin:2019da}. At last, let us consider again two square-lattice Ising models with spin variables $\sigma$ and $\tau$, now coupled via a term of the form $\sigma \tau$: is the phase with composite order parameter
  $\langle \sigma \tau \rangle$ present or not? As discussed in literature for the bilayer configurations and reviewed below, we expect that such phase should {\emph{not}} exist.
 Since the phase diagram of the 2D Ashkin-Teller model and of some its
variations can be determined analytically~\cite{Baxter:2007,Delfino04}, and similarly the Ising model is a classical workhorse of statistical mechanics~\cite{Cardy96,Mussardo10}, they
  provide an ideal benchmark to look for composite order parameters
in un unsupervised way. One could can ask whether and what new composite order parameters emerge in multilayer configurations, such as the trilayer one.
  Although in the two-variable (or, in our language, bilayer) Ashkin-Teller model the composite order parameter can be easily recognized, a more complex spin model with several layers, with both
short- and long-range interlayer couplings, could be much more challenging to be addressed with simple physical considerations. Many, possibly competing, composite order parameters may be present and
determining the one which actually breaks the symmetry and generates a novel phase is a highly non-trivial task. From this point a view, an unsupervised approach able to correctly reproduce the phase diagram of layered models, regardless of the nature of underlying order parameters, is highly desirable.

Our second motivation is that layered models emerge in a wide range of physical situations.
Among them, the bilayer structure in which two two-dimensional 
systems are coupled has been studied in a number of cases, ranging from
graphene~\cite{Novoselov04} to ultracold dipolar gases~\cite{Baranov12}.
Another major example is provided by layered supercondutors, that can occur naturally or be artificially created.
Among the former class, of primary importance are compounds of
transition-metal dichalcogenides layers intercalated with organic molecules~\cite{Gamble70} and
cuprates~\cite{Tinkham96}. Examples of artificial structures are alternating layers of
graphite and alkali metals~\cite{Hannay65} or samples with layers of
different metals~\cite{Ruggiero80}. Neutral layered superfluids can be engineered 
with quantum gases by using a deep optical lattice in one spatial direction with ultracold 
fermions~\cite{Iazzi12} or bosons~\cite{Cazalilla07}. It is therefore important to develop general approaches capable of dealing with coupled interacting
  systems. In particular, given the importance of layered physical systems and their ubiquitous presence in a variety of contexts, one may think for instance of layered superconductors, a general approach to individuate their phase diagram -- once that one is able to study the uncoupled counterpart -- would provide an important tool of investigation.

Finally, our last motivation is purely methodological and inherent to machine learning. Indeed, in layered models one has a certain degree of arbitrarity in the way the MC data to be analyzed are fed to the neural networks, \emph{e.g.}~one can provide the data in each layer separately, or retaining their spatial structure
such as columns and ordering them correspondingly. As an example, in the Ashkin-Teller model one can provide numerical algorithms either with all the $\sigma_i$'s and then all
the $\tau_i$'s, or the pairs $(\sigma_i,\tau_i)$ according to the index $i$ labeling the position of the spins in the layers.
This arbitrarity also reflects itself in the fact that an order parameter which can be clearly identified with a choice can be non-trivial,
or ``hidden'', with another choice. A special class of hidden order parameters are composite order parameters, i.e.~parameters defined across multiple layers of a layered system. To use again the Ashkin-Teller model as an example,
the order parameter $\langle \sigma \tau \rangle$ is immediately identified
when the choice of the pairs $(\sigma_i,\tau_i)$ is done, but not when the is provided layer by layer.
Therefore a natural question is how to identify phase transitions in coupled or layered models driven by order parameters which may be hidden by the
codification of the data to be provided to the machine learning algorithm. 

\section{Machine learning phase transitions in classical spin models}

Let us consider a general case of a spin system whose Hamiltonian is defined by two parameters, $J$ and $K$.
We aim to devise a procedure to depict the phase diagram in the $K-J$ plane. To this extent we discretize a
portion of the $K-J$ plane on a grid with  steps $\Delta J$ and $\Delta K$. For each point on the grid we
generate a number of uncorrelated MC snapshots using standard algorithms \cite{Swendsen:1987eq,Salas:1996ke,Wolff:1989iy}.
Unless otherwise specified we shall work on a $32 \times 32 \times N_l$ square lattice, $N_l$ being the number of layers
to be specified later, and we shall generate a number of 600 snapshots for each point in the phase diagram. Periodic boundary conditions are used on each layer throughout all the simulations.

The training of the convolutional neural network attempts at learning to distinguish
snapshots belonging to the two different points, $(J_1,K_1)$ and $(J_2,K_2)$, in the phase diagram. Intuitively, when this training fails,
the two points present nearly identical features, thus belonging to the same phase. On the other hand,
if it succeeds, the two points should belong to two different phases. In order to carry out this plan,
at first, we divide the data in a standard way, taking 80\% of snapshots from each of the two points as training data,
while keeping the other 20\% as validation data. Then, we train the network on the training data and
quantify the classification accuracy on the validation set as the fraction $\varphi$ of correctly labeled examples
from the validation set. Based on that, we introduce the following quasidistance
\footnote{We use the term 'quasidistance' since it does not respect triangular inequality. However, this fact plays no role as far as all the applications in the present paper are concerned.} between the two phase diagram points $(J_1,K_1)$ and $(J_2,K_2)$:
\begin{equation}
d((J_1,K_1),(J_2,K_2))=2(\varphi-0.5)\Theta(\varphi-0.5)\; ,
\label{eq:distances}
\end{equation}
where $\Theta(x)$ is the Heavyside step function, preventing $d$ from assuming negative values.
Then perfect discrimination $\varphi=1$ (signaling different phases) corresponds to $d=1$, while perfect confusion
$\varphi=0.5$ (signaling the same phase) corresponds to $d=0$.

We feed the raw Monte Carlo snapshots directly to the convolutional neural network, with spin down encoded as 0 and spin up encoded as 1, no preprocessing applied. The network architecture is optimized for the task of classifying two phases:  after convolutional and fully connected layers the final layer consists of two softmax output neurons outputting the labels. The convolutional filters span both layers, which is the feature enabling the network to learn composite order parameters. Hence, both layers are simultaneously fed into the network. Further technical details on the network architecture and training can be found in the Appendix.

\begin{figure}
\centering
\includegraphics[width=\columnwidth]{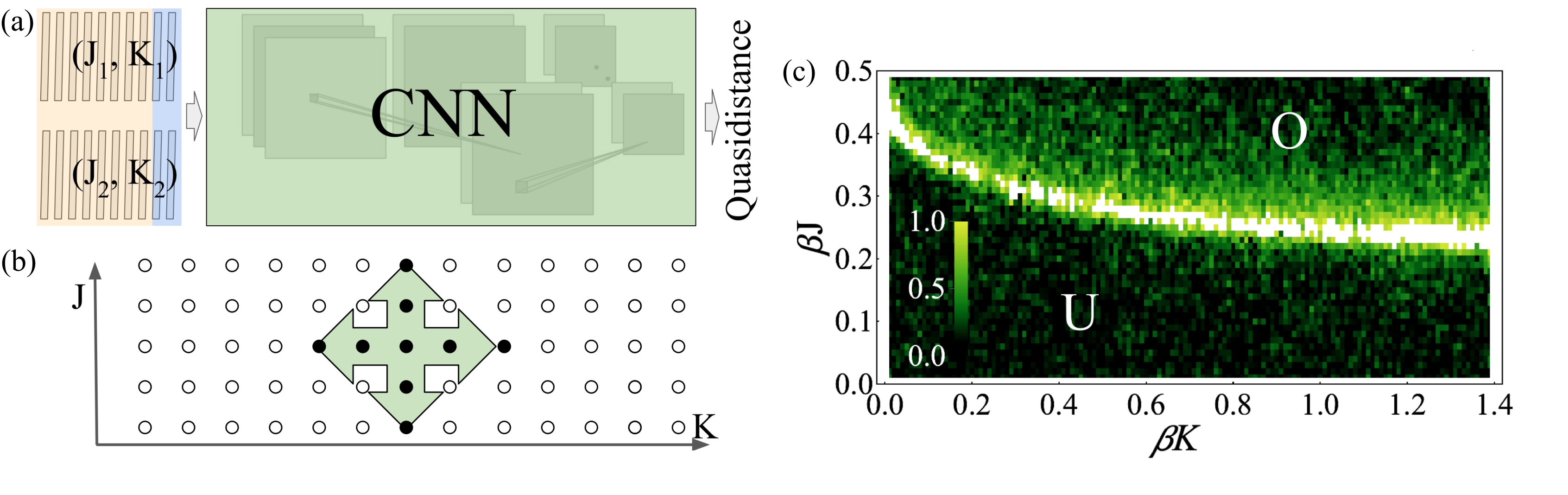}
\caption{\label{fig:visualization} An overview of the proposed method.
  (a) The convolutional neural network is able to assess quasidistance which determines whether the phase diagram points $(J_1,K_1)$ and $(J_2,K_2)$ belong to the same or different phase. This is done by attempting to learn to distinguish between individual Monte Carlo snapshots with orange and blue visualizing training and validation MC snapshots, respectively, see main text. (b) Using distances between first, second and third neighbors, one can evaluate the Laplacian across the phase diagram.
  (c) Large values of the Laplacian signal the presence of phase transition. Plotting the Laplacian reconstructs the phase diagram, which is now parametrized by the dimensionless combinations $\beta J$ and $\beta K$, with $\beta$ the inverse temperature. Here we show the reconstructed phase diagram for the square-lattice Ising bilayer model with the transition between ordered (O) and unordered (U) phases, see main text.}
\end{figure}

At last, we make use of the distances defined in Eq.~(\ref{eq:distances}) to construct a
field $u(J,K)$ defined on the phase diagram through its finite-difference lattice gradient
\begin{equation}
\nabla u(J,K) = \begin{pmatrix}
(u(J+\Delta J,K) - u (J,K)) / \Delta J \\
(u(J,K+\Delta K) - u (J,K)) / \Delta K
\end{pmatrix} 
 \equiv \begin{pmatrix}
d((J+\Delta J,K),(J,K)) / \Delta J \\
d((J,K+\Delta K),(J,K)) / \Delta K
\end{pmatrix} \; .
\end{equation}
Clearly $\nabla u$ will be constant in regions of the phase diagram belonging to the same phase, since we expect that
the difficulty of telling first neighbors apart should be uniformly quite high. On the other hand, we expect the
value of $\nabla u$ to abruptly change in the vicinity of a phase transition, suggesting that the phase diagram
should be naturally characterized by looking at the finite-difference lattice Laplacian
\begin{equation}
\nabla^2 u (J,K) \approx \frac{1}{(\Delta J)^2} \sum_{i = 0}^{n} (-1)^i \ \binom{n}{i} \ u (J + (n/2 - i) \Delta J), K) + 
\frac{1}{(\Delta K)^2} \sum_{i = 0}^{n} (-1)^i \ \binom{n}{i} \ u (J, K+ (n/2 - i) \Delta K)) \; ,
\label{eq:nablau}
\end{equation}
with the $n=2$, $n=3$ and $n=4$ cases corresponding to a 5-point,
9-point or 13-point stencil, respectively. The stencil includes $(n-1)$ nearest neighbors in the $J$ and $K$ directions. 
We stress that the summations can be rearranged so that they involve only
differences of the $u$ field evaluated between first, second and third neighbors, that can in turn be expressed in terms of
the quasidistance $d$. From the discussion above, it is clear that a sudden rise in the value of $\nabla^2 u$
means that the CNN can distinguish with increased precision arbitrarily close points in the phase diagram,
thus signaling a phase transition. We anticipate that including high-order finite-differences besides the obvious
5-point stencil taking into account first-neighbors stencil considerably increases the quality of the reconstructed phase diagram.
This point will be analyzed in detail later. Moreover, using the stencil as opposed to always just comparing two neighboring points of the phase diagram immunizes the algorithm in the case of very dense grid. In such a case, it would be progressively difficult to find neighboring points belonging to different phases. With our approach, we are assured that using a large enough stencil will circumvent this problem for any grid density.

Calculation of $\nabla^2 u (J,K)$ for the entire phase diagram is by far the most time-consuming step of the algorithm. Using $N$ nearest-neighbours, i.e. $\left[4(N-1)+1\right]$-point stencil, it requires $M\cdot 4(N-1)$ calculations of the quasidistance. There, 
\begin{equation}
M=
\frac{J_{\textnormal{max}}-J_\textnormal{min}}{\Delta J}
\cdot
\frac{K_{\textnormal{max}}-K_\textnormal{min}}{\Delta K}
\end{equation}
is the total number of discretized $(J,K)$ pairs in the phase diagram.

In conclusion of the present Section, we compare our scheme with other related approaches.
As opposed to other machine learning schemes, in the present work we do not need the evaluation of any
observable quantity to establish a distance~\cite{Broecker:2017vva}, rather directly
relying on the MC snapshots. Moreover, as opposed to other approaches~\cite{Broecker:2017vv} the scheme we introduce in
this paper fully  takes advantage of the two-dimensional nature of a two-parameter phase diagram,
as the local information is reconstructed by taking into account neighbours in all directions.
Extensions to three- or higher-dimensional phase diagrams are straightforward~\cite{Thampi:2013ij}.
Finally, our approach requires only the evaluation of a fixed number of neighbors for each point in the phase diagram,
ensuring that the computational effort required for training scales linearly with the number of points in the discretized phase diagram.

\section{Multilayer Ising models}

We now use the framework described in the previous Section to characterize the phase diagram of different coupled spin models.

Let us start from a bilayer Ising system,
described by the following Hamiltonian with a quadratic coupling term (sometimes referred to as
the Yukawa coupling):
\begin{equation}
H_\text{bilayer}=-J \sum_{\langle i j \rangle} \sigma_i \sigma_j -J \sum_{\langle i j \rangle} \tau_i \tau_j 
-K \sum_i \sigma_i \tau_i\; ,
\label{eq:hbilayer}
\end{equation}
where $\sigma_i, \tau_i = \pm 1$ are
Ising variables on a two-dimensional square lattices, whose sites are denoted by the indices $i,j$. The sums
in Eq.~(\ref{eq:hbilayer}) are over nearest-neighbor sites. When $K=0$, the system reduces to two uncoupled Ising models, having a
phase transition at the Onsager critical point $(\beta J)_c = \ln ( 1 + \sqrt{2} ) /2$~\cite{Onsager,Mussardo10},
$\beta$ being the inverse temperature. This critical point
is shifted by the presence of a finite interlayer coupling $K$. The resulting Ising critical line separating
the paramagnetic and ferromagnetic phases as a function of $K$ has been studied in the literature~\cite{Oitmaa_1975,Hansen1993,BROWER_95}. It is clear that
the bilayer system~(\ref{eq:hbilayer}) is the classical counterpart of two coupled quantum Ising chains in a transverse field,
a system that has been studied both in relation to its spectrum, phase transitions and possibility to determine an integrable
line in the space of parameters~\cite{DELFINO1998675,FABRIZIO2000647,Tsvelik,PhysRevLett.102.097203}. The classical bilayer system and the quantum coupled chains can be also related to each other by an exact mapping.

From our point of view the model described by Eq.\,\eqref{eq:hbilayer} is an excellent starting point for our investigations, especially in order to check the existence of a  composite order parameter and its relation to the phase diagram.
It is now natural to parametrize the phase diagram in term of the dimensionless combinations $\beta J$ and $\beta K$,
discretizing it for values of $\beta J \in [0,0.5]$ and $\beta K \in [0,1.4]$,
with discretization steps $\Delta \beta J = \Delta \beta K = 0.01$. We then apply the phase diagram reconstruction procedure
described in the previous Section to precisely
determine the phase boundaries in the $\beta K$-$\beta J$ phase diagram, shown in Fig.~\ref{fig:visualization}(c).
The phase transition occurs at $(\beta J)_c \approx 0.44$ in the uncoupled $\beta K = 0$ case, in agreement with analytical results~\cite{Onsager,Mussardo10}. The errors of our method on the determination of transition points are discussed in Appendix~\ref{app:err}. Then the critical
temperature gradually decreases to the strong-coupling critical temperature
$(\beta J)'_c = (\beta J)_c / 2$. The width of the peak is essentially due to the the finite-size ($32 \times 32 \times 2$)
of the lattice used for Monte Carlo simulations, whose snapshots we feed to the neural network. The result is that it appears
that only two phases are found, with order parameter $\langle \sigma \rangle=\langle \tau \rangle$. From our treatment of data we cannot determine the behavior
of the order parameter inside the two phases, whose study would be an interesting future continuation of the present results.

Next, we consider a trilayer system, whose Hamiltonian is a natural extension of the one of Eq.~(\ref{eq:hbilayer}):
\begin{align}
H_\text{trilayer}=-J \sum_{\langle i j \rangle} \sigma_i \sigma_j -J \sum_{\langle i j \rangle} \tau_i \tau_j - J \sum_{\langle i j \rangle} \upsilon_i \upsilon_j  -K \sum_i \sigma_i \tau_i -K \sum_i \tau_i \upsilon_i\; ,
\label{trilayer}
\end{align}
and the new variable $\upsilon_i$ is also an
Ising spin. This is the first non-trivial example, and of course representative
of properties of the multilayer Ising model with Yukawa coupling. The central
natural question is whether a composite order parameter emerges. Moreover the model of Eq.~\eqref{trilayer} is interesting
since it paves the way to the investigation of the $N$ layers case, which shall be trivial
with the method  presented here. Indeed the $N$ layer case may serve to investigate
how the (three-dimensional) limit of infinite layers is retrieved, an issue in the context of layered models, see
\emph{e.g.}~Ref.~\cite{Schneider98}.

\begin{figure}
\centering
\includegraphics[width=0.5\columnwidth]{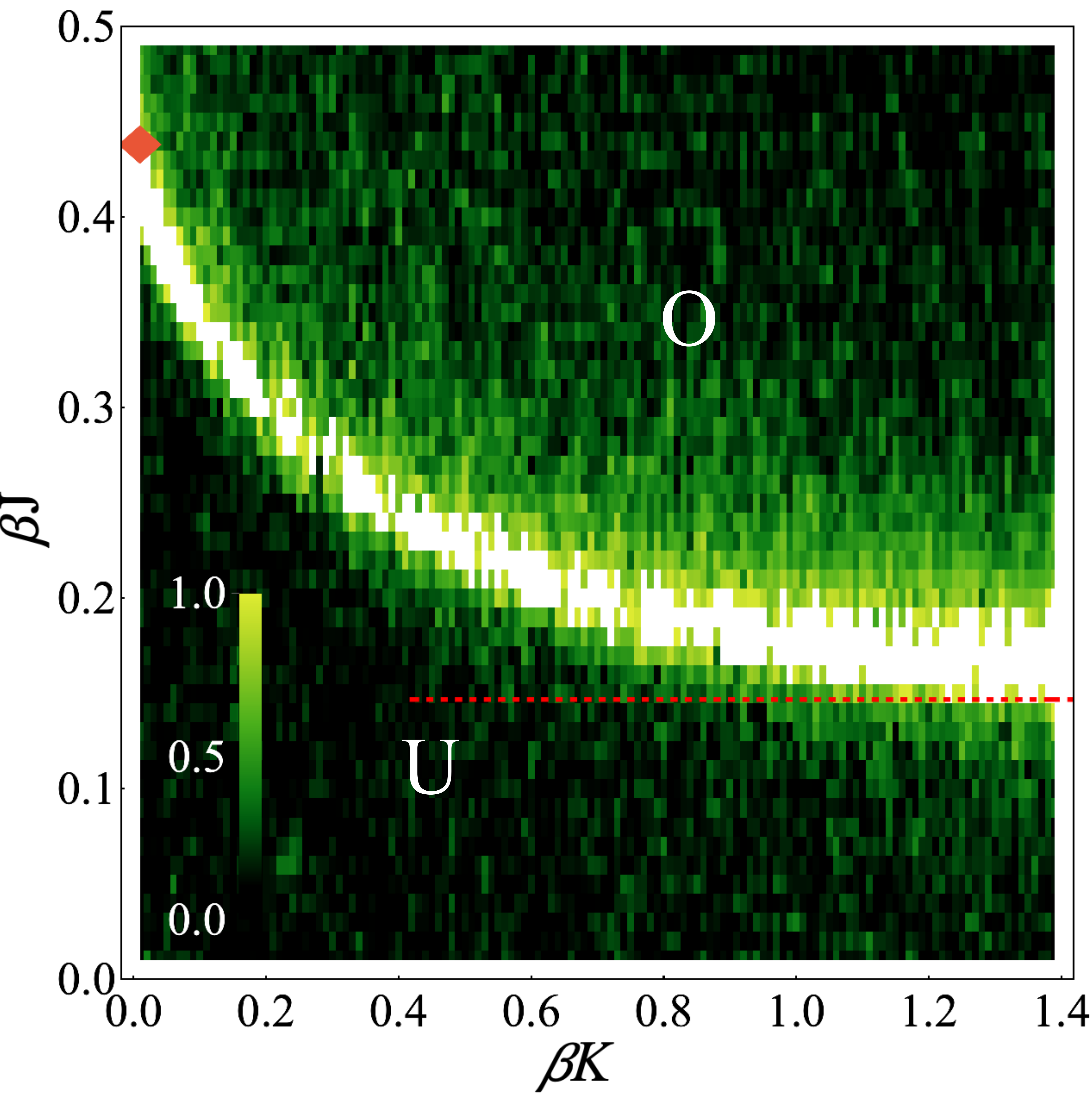}
\caption{\label{fig:trilayer} Reconstructed phase diagram for the square-lattice Ising trilayer model,
  showing a phase transition between an unordered, high-temperature phase (U) to an ordered, low-temperature phase (O).
  Note that as the interlayer interaction $\beta K$ is increased, the critical temperature decreases from the analytic limit
  $(\beta J)_c = \ln ( 1 + \sqrt{2} ) /2 \approx 0.44$, marked by a red diamond, to the strong-interlayer-coupling limit
  $\beta K \to \infty$ where $(\beta J)'_c = (\beta J)_c / 3$, marked by a red, dashed line.}
\end{figure}

The investigation of the model described by Eq.\,\eqref{trilayer} follows the same line as the one of the bilayer case, we are able to reconstruct the phase diagram as shown in Fig.~\ref{fig:trilayer},
recovering that strong-interlayer-coupling critical temperature that in this case is $(\beta J)''_c = (\beta J)_c / 3$,
marked by a red dashed line. The main result exhibited in Fig.~\ref{fig:trilayer} is that no composite order parameter appears even for the trilayer case. Therefore, our technique has been able to correctly recover the phase diagram of the bilayer Ising model, where we do not expect any additional order to appear\,\cite{Smiseth:2005hh,Sellin:2016kk}, while it also predicts the same picture for the trilayer case, for which no previous expectation exist up to our knowledge. The generalization to the $N$-layer case shall be straightforward, but more numerically demanding, while based on the present results no additional phases are expected to appear. Therefore,
in the following we are going to investigate a different case where a composite order parameter appears by construction.

\section{Reconstructing composite order parameters: the Ashkin-Teller model}

We now turn to the square-lattice Ashkin-Teller model, described by the following Hamiltonian
\begin{equation}
H_\text{AT}=-J \sum_{\langle i j \rangle} \sigma_i \sigma_j -J \sum_{\langle i j \rangle} \tau_i \tau_j 
-K \sum_{\langle i j \rangle} \sigma_i \sigma_j \tau_i \tau_j
\label{AT}
\end{equation}
with $\sigma_i, \tau_i=\pm 1$. Compared to Hamiltonians~(\ref{eq:hbilayer})-(\ref{trilayer}) one sees
that the coupling is now quartic in spins. Since in the Ising model there are only two scaling fields relevant
in renormalization group sense~\cite{Cardy96,Mussardo10}, the magnetization and the energy, one sees that in the models~(\ref{eq:hbilayer}) and~(\ref{AT}) one has basically the two natural ways of having respectively magnetization-magnetization and energy-energy couplings, higher order coupling terms being irrelevant. The Ashkin-Teller model is also related
to the four state planar Potts model, and several variations of it, also in three dimensions, have been
investigated~\cite{RevModPhys.54.235}. 

The Ashkin-Teller model features a rich phase diagram, and remarkably in two dimensions can be studied
analytically~\cite{Baxter:2007,Delfino04}. Here we consider the case of ferromagnetic couplings, $J, K \geq 0$.
It is known that three different phases exist~\cite{Baxter:2007}. Besides an ordered phase, denoted by I,
characterized by $\langle \sigma \rangle \neq 0 \neq \langle \tau \rangle$ and a disordered phase, II,
characterized by $\langle \sigma \rangle = \langle \tau \rangle = 0$ one also finds the peculiar phase III
in which the single spins $\sigma$ and $\tau$ are disordered, whereas a composite order parameter given by their product
is ferromagnetically ordered, i.e. $\langle \sigma \tau \rangle \neq 0$.

\begin{figure}
\centering
\includegraphics[width=0.5\columnwidth]{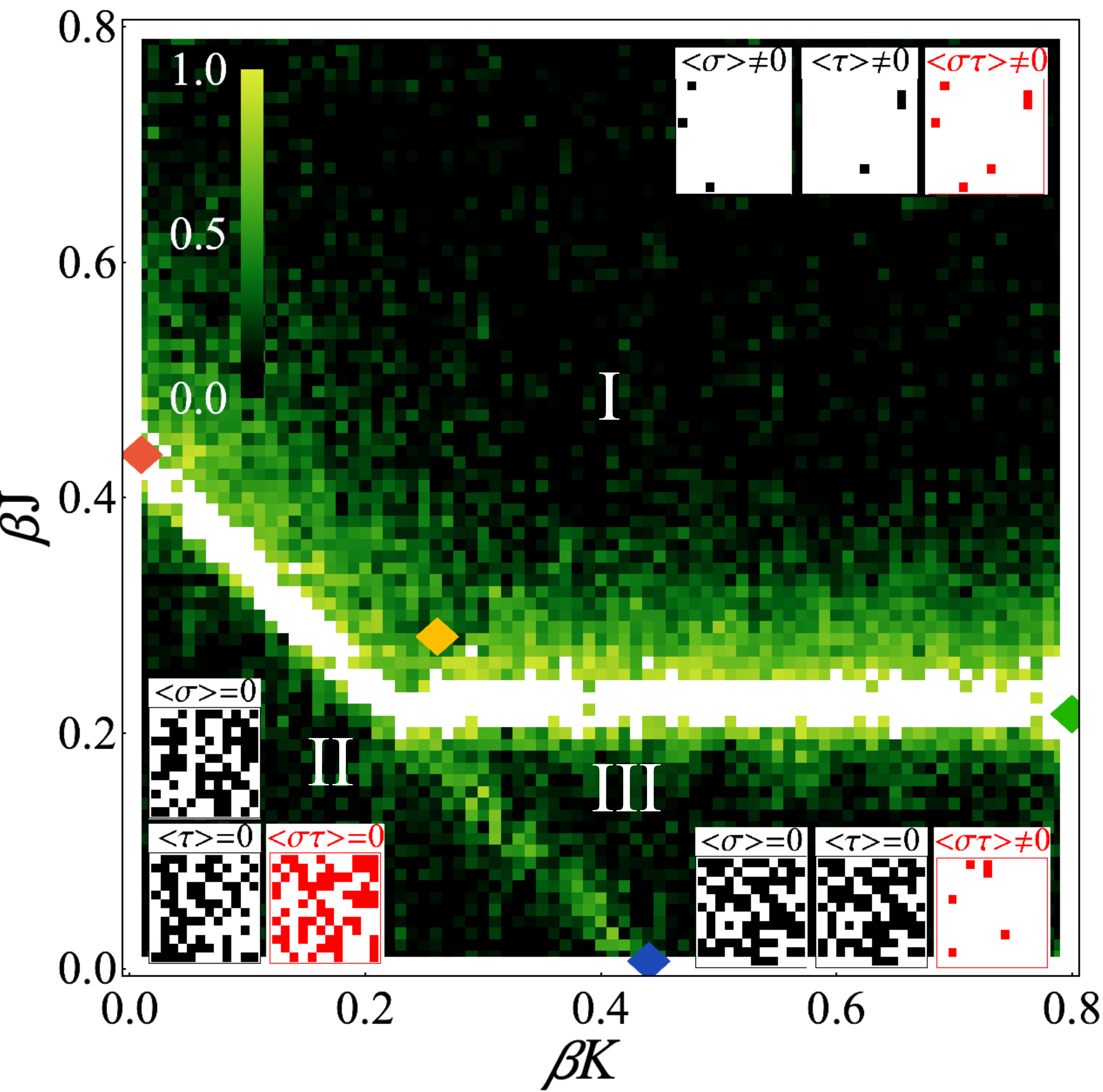}
\caption{\label{fig:ashkinteller} Reconstructed phase diagram for the square-lattice Ashkin-Teller model:
  the red, blue yellow and green diamonds show analytically-determined phase transition points, see main text.
  Our approach identifies three phases, in agreement with the theory of the Ashkin-Teller model. The insets show
  representative configurations of the $\sigma$, $\tau$ spins and of the `composite' spin $\sigma \tau$, for each phase:
  note that the transition between phase II and phase III does not correspond to any apparent difference in the
  $\sigma$ and $\tau$ layers that we feed to the CNN. We stress that the $\sigma \tau$ `composite' variable, marked in red,
  is not fed to the CNN.}
\end{figure}

Whereas the previous investigation of Ising-like models makes us confident that the ML procedure we have introduced is
able to correctly characterize the transition between phase I and phase II, it is not \textit{a priori}
clear that phase III can be correctly identified. As shown in the small inset of Fig.~\ref{fig:ashkinteller},
MC snapshots show disordered spins both in phase II and in phase III, the transition being determined by the
$\sigma \tau$ composite variable, that we do not directly feed to the CNN.
In order to learn the existence of the II-III phase transition the CNN must learn to reconstruct the composite order parameter.
We find that our framework successfully performs this task, owing to the convolutional filters which are convolved in 2D spanning across the layers and are able to learn even elusive interlayer correlations.

The reconstructed phase diagram of Fig.~\ref{fig:ashkinteller} shows that indeed our approach is able
to correctly learn the phase transitions in the ferromagnetic Ashkin-Teller model. Whereas the transition
line corresponding to the magnetization of $\sigma$ and $\tau$, as separated variables, corresponds to a prominent peak,
whose width is essentially determined by finite-size effects, the line corresponding to the magnetization of the
composite $\sigma \tau$ order parameter corresponds to a smaller peak, displaying that the characterization of this
transition line is more demanding to the CNN, but still possible.

We can compare the obtained phase diagram we obtain with some exact results.
In the $K \to 0$ the model reduces to a square-lattice Ising model with coupling constant $J$, with critical temperature
$(\beta J)_c = \ln ( 1 + \sqrt{2} ) /2 \approx 0.44$~\cite{Onsager,Mussardo10},
whereas in the $K \to \infty$ limit the model reduces to a square-lattice Ising model with coupling constant $2J$ and
critical temperature $(\beta J)'_c = \ln ( 1 + \sqrt{2} ) / 4 \approx 0.22$. Finally for $J=0$ the system again
undergoes an Ising-like phase transition for the composite order parameter, at $(\beta K)_c = \ln ( 1 + \sqrt{2} ) /2 \approx 0.44$.
These three points are marked by a red, green and blue diamond, respectively, in the phase diagram of Fig.~\ref{fig:ashkinteller},
showing an excellent agreement between the analytical results and the reconstructed phase diagram, even in the latter case
when the composite order parameter $\sigma \tau$ drives the transition. Finally, the yellow diamond marks the bifurcation point as determined analytically in Ref.~\cite{Baxter:2007}; we attribute the difference with respect to the bifurcation point in our reconstructed phase diagram to finite size effects. We also mention that the critical lines separating the different phases are retrieved with a precision up to $\sim 20\%-30\%$, except for vanishing $\beta J$. Again we attribute this to finite size effects; proceeding as extensively discussed in the literature~\cite{Carleo:2019us} one could obtain a quantitative agreement on the location of the critical lines. Here, our emphasis is on the possibility of retrieving the phases with composite order parameters and to ascertain their existence, as we also did for the trilayer Ising model.

\begin{figure}
\centering
\includegraphics[width=0.5\columnwidth]{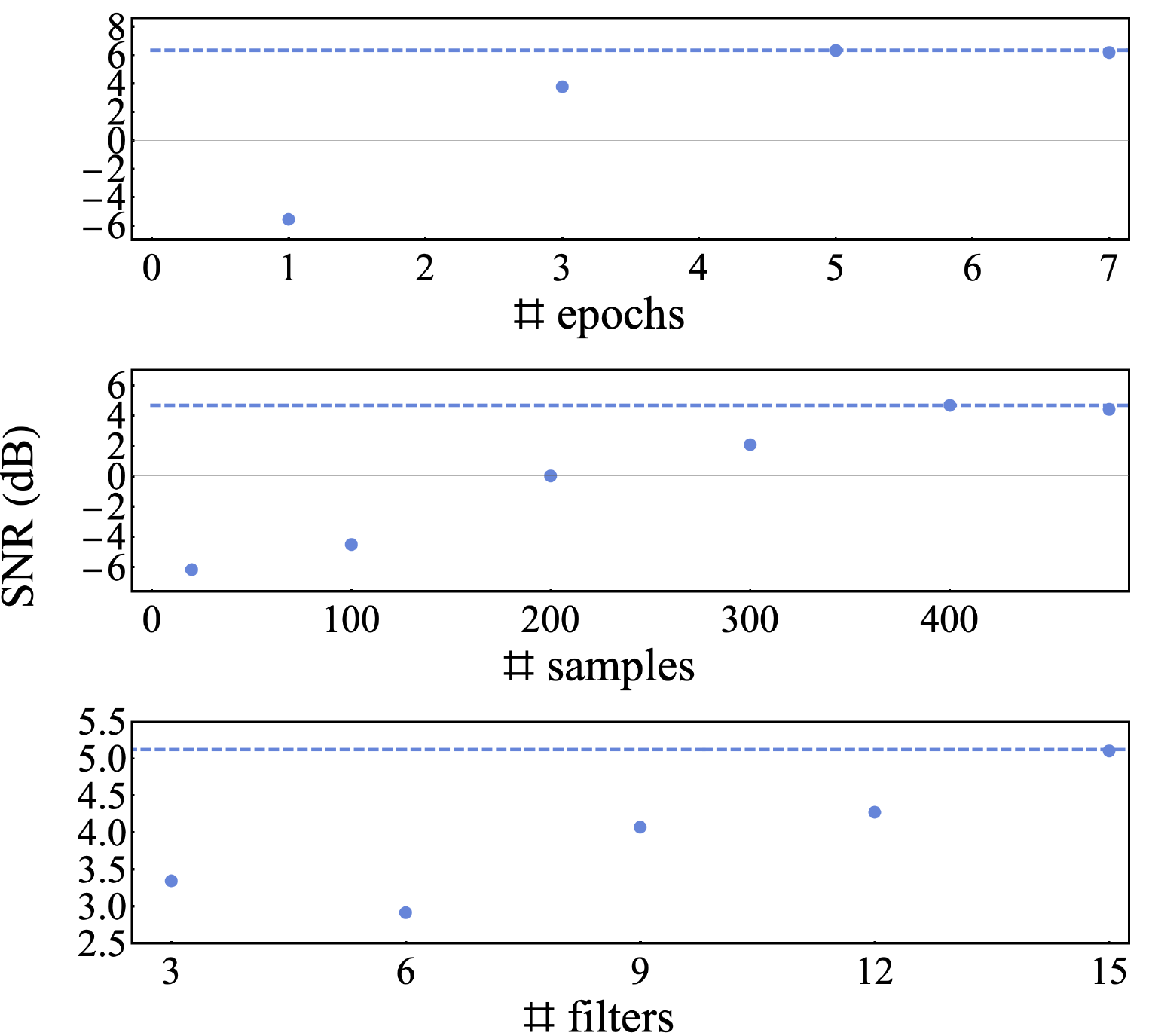}
\caption{\label{fig:scalings} Signal-to-noise ratio for the Ising bilayer as a function of the number
  of epochs (upper panel), of the number of samples in the training set (middle panel) and of the number of
  convolutional filters (lower panel). The dashed lines guide the eye towards the highest attainable signal-to-noise ratio in each dataset.}
\end{figure}

\section{Scaling properties and robustness of~the~approach}
Our results show that with the network and learning parameters that we used we were able to obtain a phase diagram of
quality high enough to visually identify different phases. In addition, in this Section we characterize our method
by quantifying signal to noise ratio (SNR) and studying its behavior when essential parameters are changed.
We define the SNR as
\beq
\text{SNR} \equiv \log_{10} \left( \frac{\frac{1}{N} \sum_i (x_i - \nu)^2}{\nu^2} \right)\; ,
\eeq
$x_i$ being the values of the $\nabla^2 u$ field of Eq.~(\ref{eq:nablau}),
the summation extending over a region containing $N$ values, $\nu$ being the `noise', i.e. the average value of
$\nabla^2 u$ in a subset of the region far away from a phase transition. We evaluate the SNR for
the Ising bilayer on a strip centered on $\beta K=1.1$, exhibiting a sharp phase transition at $\beta J \approx 0.26$ as
clear from Fig.~\ref{fig:visualization}. At first, we vary the number of training epochs, observing that the
SNR is rapidly increasing before reaching a maximum value at around 5 epochs of training.
This indicates that further training brings no benefit while providing a risk of overfitting,
justifying our early-stopping approach. Secondly, we vary the number of samples in the training set,
showing a rapid increase in the SNR before reaching a plateau at about 400 samples, justifying our choice of using a slightly
larger number (600) of samples in the training set. Lastly, we vary the number of convolutional filters in the CNN.
Again, the general upwards trend shows that a larger number of convolutional filters helps in enhancing the quality
of the reconstructed phase diagrams. However, we stress that in this latter case the SNR is quite high
in the whole parameter region we consider. The lowest number of convolutional filters
we consider (3) is already enough for achieving a good reconstruction of the phase diagram and a large SNR value.
These analyses are shown in Fig.~\ref{fig:scalings}.

We have also analysed how the reconstructed phase transition is affected by the dimension of the stencil in Eq.~(\ref{eq:nablau}).
Using a 5-point, 9-point or 13-point stencil we have obtained SNR values of $-1.36 \ \mathrm{dB}$, $0.38 \ \mathrm{dB}$ and
$3.88 \ \mathrm{dB}$, respectively. This confirms that the approach we are introducing takes indeed
great advantage from the two-dimensional structure of the phase diagram, and information from second- and
third-nearest neighbors is being used to sharply characterize the phase transition.

\section{Conclusions}

Thus, as shown for layered spin models such as
the multilayer Ising and Ashkin-Teller models, our work demonstrates that ML approaches are able to learn the order parameter driving a phase transition in layered models, also
when this parameter is not immediately apparent from the snapshots without preprocessing. This is directly possible due to the convolutional filters which are, without any a priori knowledge, capable of learning even involved algebraic operations that uncover the order parameters from the data.
This paves the way to the use of ML approaches to investigate the properties of systems of increasing complexity
and to characterizing phases of matter described by multiple, possibly non-local order parameters,
the universal approximation theorem~\cite{Hornik:1991kg} ensuring that a neural network can, at least in principle,
learn to recognize arbitrarily complex order parameters. In particular it would be very interesting to study the multilayer Ising model
with a number of layers increasing, the three-dimensional Ashkin-Teller and the trilayer Ahkin-Teller in two dimensions,
which can be studied with the techniques introduced in the paper. Non-local couplings among the layers could be added,
which would lead to non-local, more composite, operators. These results should be compared with the identification of hidden order
done using non-ML techniques~\cite{Martiniani:2019go}. Also, the present approach may be used for other cases
in which the identification of the order parameters is not straightforward~\cite{Riggs:2015jh,Lee:2017kk,Varma:2006ea}.
Even if our approach has been devised to deal with coupled spin models and can applied to different geometrical configurations, it is not clear {\emph{a priori}} that it would succeed in other more complicated cases of coupled interacting systems, such as multilayer configurations of interacting bosons and fermions or bilayer quantum Hall systems. Of course, in order to study generically coupled models one needs to have an efficient algorithm to simulate the uncoupled systems. Nevertheless, we think the present work provides a methodological basis, highlighting the effect of interlayer coupling on the macroscopic properties and phases of coupled systems.

Naturally, the approach we introduced could also be extended in the future to characterize quantum models, or classical spin models with competition between short- and long-range interactions, or more involved spin models such as the XY model, discretising the continuous degrees of freedom\cite{Lupo:2017}. We expect
that by an appropriate choice of the sizes and strides of the filter in the convolutional layer one could
characterize antiferromagnetic order parameters, non-local order parameters and exotic order parameters,
such as nematic and smectic phases. In this context, current experiments on fermionic dipolar atoms\,\cite{Lu2012,Park:2015dj} promise to open a new window in the physics of competing long-range and short-range interactions\,\cite{Baier:2018kx}, clearing the path for the comprehension of modulated phases in strongly interacting quantum systems.

The presence of spatially ordered structures is a {\emph{leitmotiv}} for long-range and layerd systems such as ultra-thin magnetic films\,\cite{Allenspach:1992ip,Kashuba:1993ki,Kashuba:1993ds}, iron-based superconductors and cuprates\,\cite{Parker:2010if,Tanatar:2015is}. The pattern structure normally depends on several experimental conditions and it produces a particularly rich phase diagram. Most of the common features occurring in stripe forming systems and modulated phases remain obscure due to the challenges posed by the complicated order parameters, which occur in these cases\,\cite{MendozaCoto:2016fm,MendozaCoto:2012dw,Barci:2008gz,Barci:2011bm}. The ML technique introduced in the present paper may serve as an essential prove to finally uncover the complexity of such phases.

Our results pave the way for fully automated study of phase diagrams of more general and complicated spin systems.
An exciting open problem lying in the realm of so-called explainable artificial intelligence (XAI)~\cite{XAI}
is whether machine learning techniques could not only learn to separate phases differing by a `hidden' order parameter,
but also identify that parameter. Another natural development of the present work is to use our
fully-unsupervised technique to learn directly from experimental data~\cite{rem2019identifying,Bohrdt:2018tr, Zhang:2018iz}. Finally,
it would be interesting to extend the results presented in this paper according to the variational procedure discussed
in Ref.~\cite{Variational}.

\ack
We thank Gesualdo Delfino, Michele Fabrizio, Piero Ferrarese, Robert Konik, Christoph Lampert and Mikhail Lemeshko for stimulating discussions at various stages of this work. W.R.~has received funding from the EU Horizon~2020 programme under the Marie Sk\l{}odowska-Curie Grant Agreement No.~665385. G.B.~acknowledges support from the Austrian Science Fund (FWF), under project No.~M2461-N27. N.D. acknowledges support by the Deutsche Forschungsgemeinschaft (DFG) via Collaborative Research Centre SFB 1225 (ISOQUANT) and under Germany's Excellence Strategy EXC-2181/1-390900948 (Heidelberg STRUCTURES Excellence Cluster).

\section*{References}
\bibliography{extracted}

\appendix

\section{Details on the architecture and on the training of the convolutional neural network}
\begin{figure}[h]
\centering
\includegraphics[width=0.5\columnwidth]{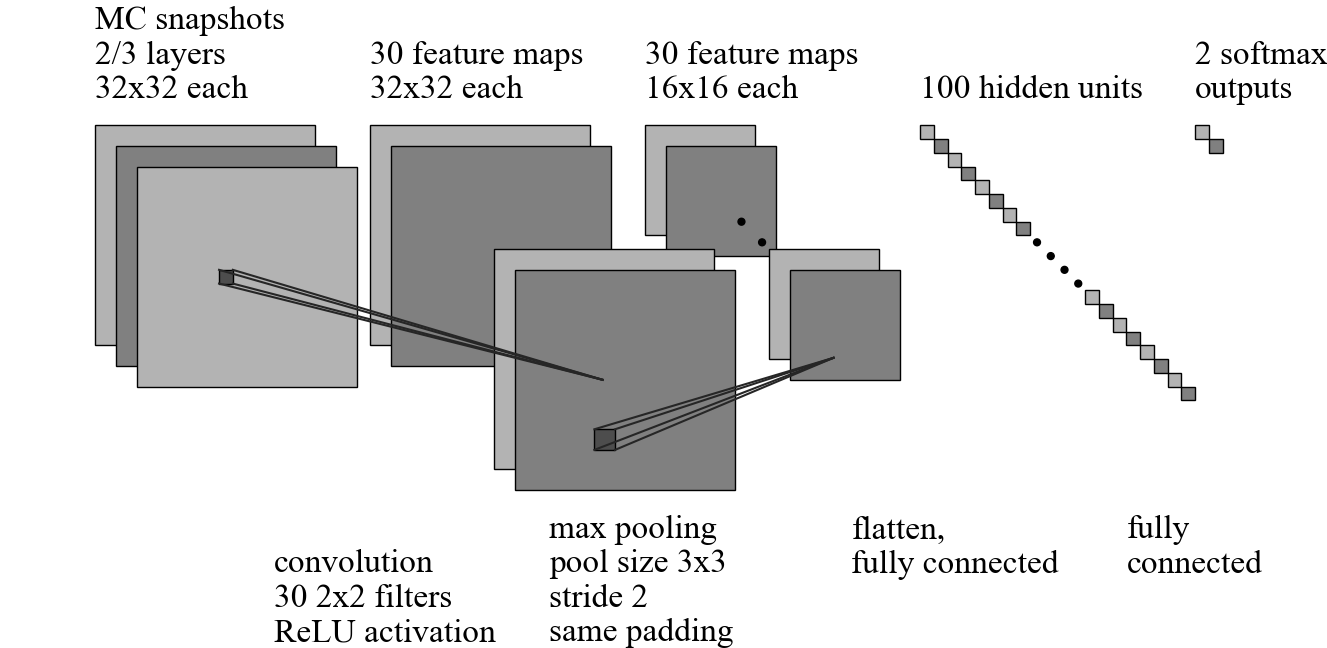}
\caption{\label{fig:convnet_architecture}
  Visualization of the convolutional neural network used. Lower labels describe the layer operations. Upper labels describe the shapes of tensors before and after each operation.}
\end{figure}

The first layer is a convolutional layer with 32 filters of size 2 by 2 and unit stride in both directions. Then the `max pooling'~\cite{MaxPooling} operation is applied with pool size 3 by 3, stride 2 in both directions and same padding. The results is then fully connected to a hidden layer with 100 neurons.
The binary classification is finally done in the output softmax layer with two neurons. Both the convolutional and
hidden fully connected layers are activated by rectified linear units (ReLU)~\cite{ReLU}. The network is visualized in Fig.~\ref{fig:convnet_architecture}.
We train the network by minimizing the cross-entropy using the Adam~\cite{Adam} adaptive optimization algorithm with 7 epochs and minibatch size 25. Such choice leads
to a fast training -- the amount of training is much lower than in computer vision applications, routinely requiring
hundreds or thousands of epochs -- as well as prevention of overfitting by early stopping, hence eliminating
the need for other measures such as dropout~\cite{Dropout}. We use the following Adam algorithm parameters:
learning rate $0.001$ and standard choices of $\beta_1=0.9$ and $\beta_2=0.999$.
We use Tensorflow~\cite{TensorFlow} for the implementation.

\section{Finite-size scaling and the error of the proposed method}
\label{app:err}
\begin{figure}[h]
\centering
\includegraphics[width=0.5\columnwidth]{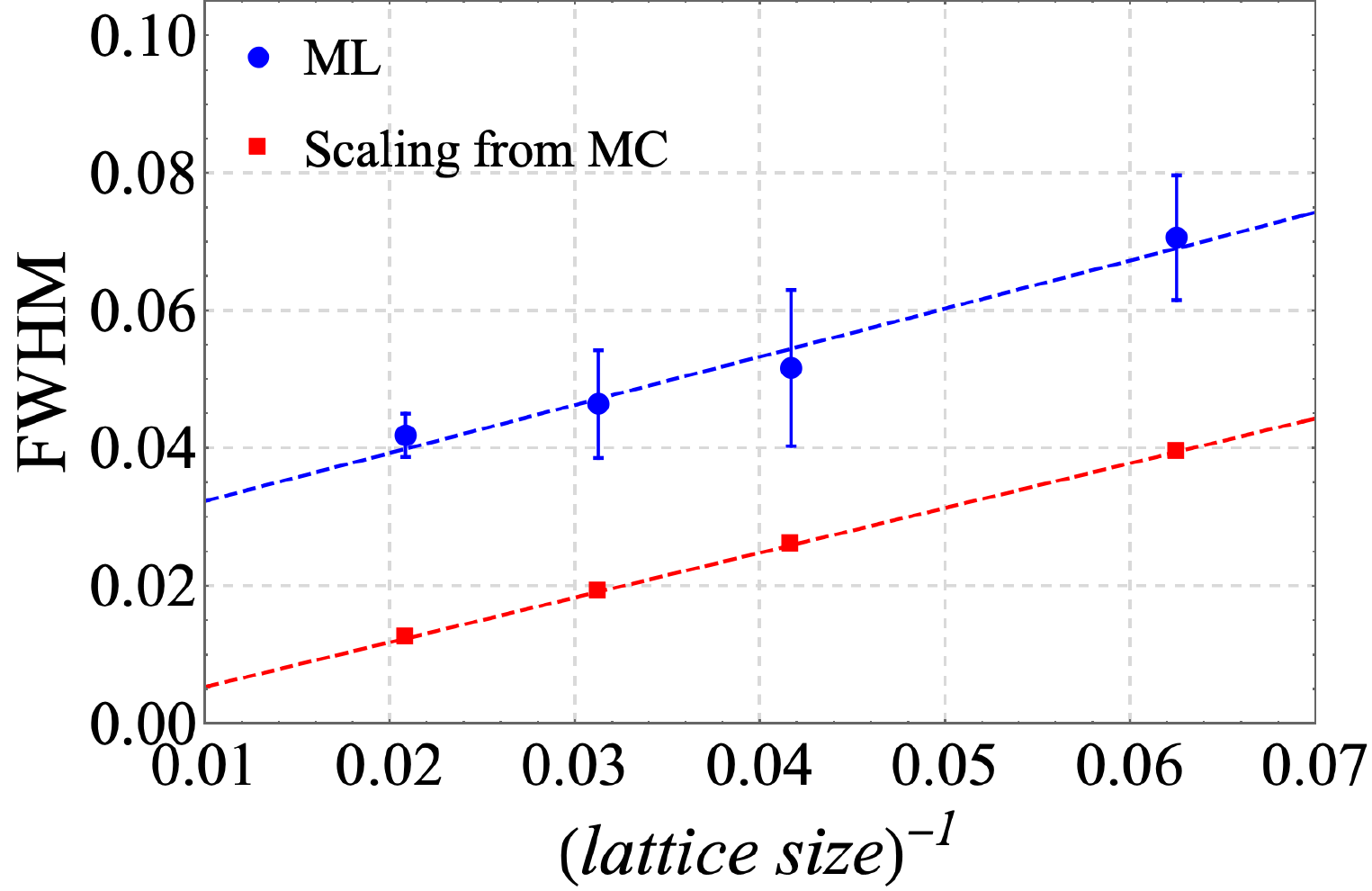}
\caption{\label{fig:errorbars}
  Full width at half maximum of the phase transition in the Ashkin-Teller model for $\beta K = 0.7$, as a function of the inverse lattice size, for four different lattice sizes ($L=16, 24, 32, 48$). The red squares show the FWHM of the peak in magnetic susceptibility in the MC data, whereas the blue squares show the FWHM of the Laplacian peak obtain from our ML approach; errors are estimated by identifying the phase transition $10$ times. The dashed lines guide the eye.}
\end{figure}
The phase transitions in the systems we consider have a certain `natural' width, due to the finite size of the lattice in the underlying Monte Carlo simulations; moreover, we expect our approach to introduce an additional width when determining the transition point. In order to verify this assumption and to investigate the accuracy of our method, we analyzed the natural width associated to the phase transition in the Ashkin-Teller model for $\beta K = 0.7$, determining it by looking at the peak of magnetic susceptibility directly from Monte Carlo simulations, and determining its width through the full width at half maximum (FWHM). We compare it with the FWHM of the Laplacian peak we reconstruct from our machine learning approach. The results are shown in Fig.~\ref{fig:errorbars}; the FWHM of both the magnetic susceptibility (red squares, the red dotted line guides the eyes) and machine learning Laplacian (blue squares, the red blue line guides the eye) obey the same $\propto 1/L$ scaling with respect to the lattice size $L$. The constant offset between the two datasets can be understood, as anticipated, as the additional error introduced of our method, due to the discretization of the parameter space, and due to some intrinsic uncertainty associated to the machine learning procedure.
\end{document}